# Capturing High-level Nondeterminism in Concurrent Programs for Practical Concurrency Model Agnostic Record & Replay

Dominik Aumayr[a], Stefan Marr[b], Sophie Kaleba[b], Elisa Gonzalez Boix[c], and Hanspeter Mössenböck[a]

a   Johannes Kepler University Linz, Austria
b   University of Kent, United Kingdom
c   Vrije Universiteit Brussel, Belgium

**Abstract**   With concurrency being integral to most software systems, developers combine high-level concurrency models in the same application to tackle each problem with appropriate abstractions. While languages and libraries offer a wide range of concurrency models, debugging support for applications that combine them has not yet gained much attention. Record & replay aids debugging by deterministically reproducing recorded bugs, but is typically designed for a single concurrency model only.

This paper proposes a practical concurrency-model-agnostic record & replay approach for multi-paradigm concurrent programs, i. e. applications that combine concurrency models. Our approach traces high-level *nondeterministic events* by using a uniform model-agnostic trace format and infrastructure. This enables ordering-based record & replay support for a wide range of concurrency models, and thereby enables debugging of applications that combine them. In addition, it allows language implementors to add new concurrency models and reuse the model-agnostic record & replay support.

We argue that a concurrency-model-agnostic record & replay is practical and enables advanced debugging support for a wide range of concurrency models. The evaluation shows that our approach is expressive and flexible enough to support record & replay of applications using threads & locks, communicating event loops, communicating sequential processes, software transactional memory and combinations of those concurrency models. For the actor model, we reach recording performance competitive with an optimized special-purpose record & replay solution. The average recording overhead on the Savina actor benchmark suite is 10 % (min. 0 %, max. 23 %). The performance for other concurrency models and combinations thereof is at a similar level.

We believe our concurrency-model-agnostic approach helps developers of applications that mix and match concurrency models. We hope that this substrate inspires new tools and languages making building and maintaining of multi-paradigm concurrent applications simpler and safer.

**ACM CCS 2012**
- **Computing methodologies** → **Concurrent programming languages**;
- **Software and its engineering** → **Software maintenance tools**;

## The Art, Science, and Engineering of Programming

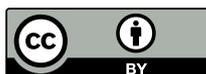

**Capturing Nondeterminism for Concurrency Model Agnostic Record & Replay**

## 1 Introduction

With concurrency being in wide use, developers started to combine the high-level concurrency models provided by languages and libraries. For example, Java and Scala complement threads & locks with libraries for actors[1] and fork/join [18]. In Scala programs, some developers use threads for efficient I/O, and actors for asynchronous communication [31]. In Go, developers frequently combine channels with shared memory synchronization to handle application requirements with suitable abstractions [33]. Such studies show that concurrent applications are *multi-paradigm*, i.e. they combine concurrency models within a single application.

Unfortunately, debugging tools have not followed this trend, making the debugging and maintenance of such applications hard. The key issue for debugging concurrent applications is their inherent nondeterminism. The order of concurrent operations can change because of timing differences. This makes the reproduction of concurrency bugs, e. g. race conditions, atomicity violations, and bad message interleavings, hard since they may manifest rarely.

Record & replay [9] assists developers by recording the nondeterminism, e. g. the outcome of races, of a specific program execution. The program can then be re-executed and diagnosed deterministically. However, existing record & replay tools are designed for a *single* concurrency model [3, 4, 19, 20, 27]. Moreover, thread-based record & replay tools, have been studied extensively [4, 9, 19, 20, 27], while high-level concurrency models have not received such treatment. Exceptions are Jardis [6], which is designed for the JavaScript event loop, and our prior work for actors [3]. In this work, we focus on capturing the nondeterminism of high-level concurrency models in multi-paradigm concurrent applications. Strategies to handle low-level shared memory can be found in the mentioned work on thread-based record & replay.

One may argue to use a thread-based tool to replay high-level concurrency models as they are often built atop the thread model. However, run-time overhead can be significantly reduced when a tool leverages the model's properties that restrict nondeterminism. For example, actors isolate state from each other, but a thread-based tool cannot leverage this isolation. It may see the same actor being executed by different threads and thus, has to record all state changes even though recording the order in which messages are executed would suffice [3].

A separate record & replay tool for each concurrency model is also insufficient. There are simply too many concurrency models and variations thereof [11]. Moreover, a record & replay tool for multi-paradigm concurrent programs also needs to capture the interactions between models to replay a program, e. g. threads interacting with actors in Scala applications.

In this paper, we propose a record & replay approach for multi-paradigm concurrent application which: (1) enables deterministic replay for high-level concurrency models (2) is practical, with a recording overhead and trace size comparable to a specialized debugging tool. Our solution records & replays high-level nondeterministic events that determine the order and outcome of the specific operations of concurrency models.

---

[1] *Akka*, Lightbend, Inc., https://akka.io/

These events are recorded in a *uniform* trace format by these specific operations, which enables support for interactions between different concurrency models. Operations are instrumented individually, which gives us the flexibility to choose a suitable and efficient approach for each operation by utilizing the concurrency model's properties.

We show our concurrency-model-agnostic record & replay by implementing support for four concurrency models and two different recording strategies. Our prototype is built on SOMns, a Newspeak [8] implementation that supports the four chosen concurrency models, and reaches performance competitive with NodeJS [23]. For the Savina benchmark suite [16], our evaluation shows tradeoffs between the recording strategies in both trace size and recording overhead. Our sender-side strategy has an average run-time overhead of 10 % (min. 0 %, max. 23 %), which is on par with a specialized record & replay for actors [3]. However, the traces are on average 24 % larger (min. −34 %, max. 111 %). In contrast, the receiver-side implementation has a higher overhead of 13.18 %, but its traces are on average only 3 % larger. For the other concurrency models, we see similar performance properties.

The contributions of this paper are:

- a *concurrency-model-agnostic record & replay approach* that can be applied to a wide range of concurrency models and combinations of them to enable e. g. deterministic debugging of multi-paradigm concurrent applications.
- a uniform and flexible trace format, capturing the nondeterministic events,
- an implementation that supports four concurrency models and two different recording strategies, demonstrating the flexibility and versatility of the approach,
- and an evaluation demonstrating that performance and trace sizes are practical and on par with a comparable system specialized for actors.

## 2 Background

As context for our work, we present a multi-paradigm application, the system in which we prototype our record & replay, and an overview of its concurrency models.

### 2.1 Multi-Paradigm Concurrent Applications

Developers have been combining concurrency models in Scala [31], Go [33] and Java applications. For instance, NetBeans, a Java IDE written in Java, uses various asynchronous event handling systems, transactional systems, as well as the concurrency abstractions provided by Java's standard library.[2]

As a concrete example of such a *multi-paradigm* concurrent application, consider figure 1 showing a system that processes sales information as a stream of JSON snippets. The sales processing system uses actors to model subsystems. Each subsystem uses a different concurrency model solving a specific problem at hand. The first subsystem, the JSON Parsing Actor, extracts `SalesData` objects from the input, and sends them to the Storage System Actor. For parsing, it uses a specialized JSON parser based on two processes communicating with channels. The first process tokenizes the input and the second extracts the relevant data, which is then returned to the actor wrapping this subsystem. The Storage System Actor interacts with a larger system that uses software transactional memory system to store the sales data and achieve reliable atomic updates of the data. The third and final subsystem, the Forecast Actor, retrieves data periodically to generate sales forecasts. These calculations are parallelized on a per-project basis using threads.

---

[2] We inspected the NetBeans codebase informally around 2014, and it has not changed significantly since then.

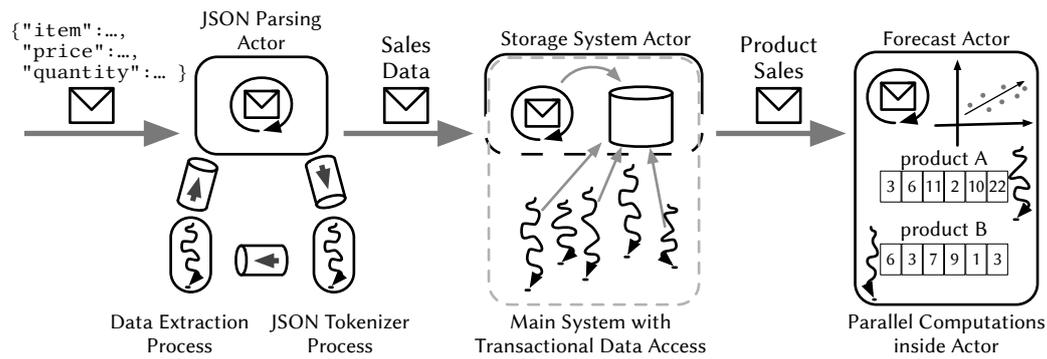

**Figure 1** A sales processing system combining different concurrency models.

We use such a sales processing system in section 6 to demonstrate that our approach is capable of recording and replaying such multi-paradigm concurrent applications.

## 2.2 Selection of Concurrency Models

In order to demonstrate the applicability of our approach, we select four concurrency models to implement our concurrency-model-agnostic record & replay: two shared memory and two message passing models. *Threads & Locks* corresponds to the widely used shared memory model and is perhaps the "standard" concurrency model. We also selected *software transactional memory* (STM) [29], which uses the notion of atomic transactions instead of critical sections to synchronize access to shared memory, making it distinctly different.

The actor model is based on message passing between concurrent entities called actors. Originally proposed by Hewitt, Bishop, and Steiger [14], many variations have since been developed [2, 11, 25, 34]. We choose the *communicating event loop* (CEL) variant, which is used in languages such as AmbientTalk [34] and E [25]. It is also the variant closest related to JavaScript's widely used event loop model. Finally, we selected the *communicating sequential processes* (CSP) model as the second model based on message passing [15]. In contrast to the actor model, the CSP model relies on processes reading from and writing to channels for message passing. We choose it for its different characteristics, and practical relevance for instance in the context of Go.

While there are many more concurrency models, the selected ones are arguably representative for a wide range of models used in today's applications (e.g. the sales processing system). Moreover, studies show that they are used in combination with each other [30, 31, 33], and thus, would benefit from better debugging support.

## 2.3 SOMns

This section presents the system in which we prototype our approach: SOMns, an implementation of the Newspeak language [8]. Newspeak is a class-based dynamically-typed language in the Smalltalk tradition. SOMns is built on top of the Truffle language-implementation framework [36, 37] and uses the GraalVM's just-in-time compilation support to reach performance competitive with NodeJS [23].

We built our work on SOMns since it supports the four selected concurrency models as part of the language. Other systems support some concurrency models directly in the language and import some others as libraries: this would make experiments more complex, because it requires different handling for recording nondeterministic events. By building on SOMns, we can directly adapt the implementation of each concurrency model as needed to realize our approach for event recording. This all makes SOMns an ideal platform for investigating multi-paradigm record & replay.

### 2.3.1 Concurrency Models, and Their Sources of Nondeterminism

While nondeterminism is inherent to most concurrency models, each one has its own features where it can originate. We now analyze where nondeterminism originates in SOMns's concrete implementation of each of the models selected in section 2.2.

**Threads & Locks (T&L)**   SOMns supports shared memory concurrency with threads and locks based on Java's semantics. Threads execute code and interact with each other through shared memory. Locks and condition variables coordinate these interactions by synchronizing the memory access to prevent data races. As outlined in section 1, we focus in this work on high-level concurrency abstractions. Thus, while we are aware that low-level memory access can lead to data races, we consider only interactions with locks and condition variables.

Details on low-level memory nondeterminism is outside the scope of this paper and can be found e.g., in related work [4, 9, 19, 20, 27].

Due to scheduling and timing differences, the order in which threads enter a critical section may vary between multiple executions of the same program. In languages such as Java and SOMns, a condition variable can only be interacted with while the associated lock is held. Thus, any nondeterminism in the interaction with condition variables is tied to the lock. Specifically, when a thread waits on a condition with await(), it releases the corresponding lock. On return from await(), the lock is reacquired. Another source of nondeterminism can be found in a variant of the await() operation that can time out, and which returns a boolean to encode whether it did.

**Communicating Event Loop (CEL) Actors**   In the *communicating event loop* (CEL) actor model [25, 34], an actor consists of a mailbox that stores received messages in order of arrival, an isolated state that can be accessed only by the actor, and an event loop that perpetually removes and processes messages from the mailbox. Like many actor implementations, SOMns uses a thread pool to execute actors rather than assigning a dedicated thread per actor. CEL actors use *promises* [25] to return results asynchronously. For instance, when a message is sent, a promise is immediately returned as a placeholder for the result. As in the E language, SOMns's promises are eventual references for the result, which can be retrieved by registering an asynchronous callback with the promise. Messages intended for the result can be sent to the promise even if the result is not yet computed. The promise stores those messages and forwards them to the result when it becomes available.

Since the CEL model isolates actors from each other, nondeterminism manifests only in the order in which an actor receives and processes messages. Unresolved promises store messages sent to them, but messages to resolved promises are sent directly to the result, and thus, are delivered to the actor mailbox. This race between the sending of messages to a promise and its resolution is another source of nondeterminism.

**Communicating Sequential Processes (CSP)** SOMns offers a CSP model with classic rendezvous semantics on channels, i. e. without buffering. As in occam-$\pi$ [35], read/write operations on channels block until another process is there to send or receive a message, respectively. In many Go programs channels without buffering are actually used more frequently than channels with buffering [12].

In addition to having rendezvous semantics, SOMns sequentializes operations on the same end of a channel. This means for instance a read needs to finish before the next one is attempted. In such CSP systems, nondeterminism manifests in the order that multiple processes read from and write to channels. Thus, CSP programs with only one reader and one writer per channel are deterministic since there is no race between readers or between writers. A program with multiple readers or writers may observe different rendezvous pairings between executions, i. e. readers may receive different messages due to the nondeterministic read/write orders.

**Software Transactional Memory (STM)** Software transactional memory [29] uses transactions to perform memory updates either completely or not at all. Only updates made by successful transactions become visible and aborted transactions do not modify the program's state. SOMns's STM is based on Renggli and Nierstrasz [28]. To ensure isolation and atomicity, transactions use a working copy of objects instead of the objects visible to the rest of the system. Thus, reads and writes are performed in isolation. When a transaction reaches its end, a global commit lock serializes transactions. A transaction then checks if the snapshots created during the transaction are still equivalent to the objects in the shared memory. If they are not, i. e. the commit of another transaction changed an object, the transaction aborts and starts over. If there are no conflicts between the snapshots and the shared memory, the transaction can commit its changes to shared memory and make them visible to the rest of the system.

In SOMns, each transaction is retried until it succeeds. Since failed transactions have no observable side effects, nondeterminism manifests in the transaction commit order. While the number of retries before a transaction succeeds is nondeterministic, this nondeterminism cannot influence the further execution of the program.

## 3 A Concurrency-Model-Agnostic Approach for Record & Replay

To realize multi-paradigm record & replay, we propose a common substrate that utilizes the commonalities of the concurrency models analyzed in section 2.3. While the semantics of the individual models differ, there are similarities in the structure of the concepts. In previous work, we presented a taxonomy of concurrency concepts for a concurrency-agnostic debugger protocol [24]. This provided inspiration for the identification of key concepts and sources of nondeterminism we employ in this work summarized in Table 1.

For each of the concurrency models, we model the execution of a program with the concept of an *activity*. Communication and synchronization operations are modeled by *interactions* between activities and *operations* on *passive entities*, i.e., objects and resources an activity operates on. These interactions and operations correspond to the sources of nondeterminism for each concurrency model. For example, the threads & locks model has *threads* as activities. Threads interact with the passive entities *conditions* and *locks*. In the CEL model, actors are the activities and interact using messages and promises.

The sources of nondeterminism we identified in section 2.3.1 correspond generally to interactions of entities that establish *happens-before* relationships [17]. For threads and locks (T&L) the sequence in which locks are acquired and whether operations timed out need to be recorded. For communicating event loops (CEL), we need to record the sequence of messages and promise resolutions. Communicating sequential processes (CSP) need to record the sequence of read/write accesses to channels. For SOMns's STM, we merely need to record the commit order.

Given these basic notions, we assume that programs are deterministic in all other parts.[3] This means that for a given input, the sequential behavior of an activity is executed fully deterministically, and all nondeterminism is modeled explicitly via the identified operations. Hence, a record & replay system only needs to capture the order and outcome of these nondeterministic operations to enable deterministic reproduction of an execution during replay. Assuming all operations with nondeterministic results are correctly identified, this eliminates all nondeterminism in replay.

Our multi-paradigm record & replay uses the high-level properties of each concurrency model and its implementation to keep tracing overhead low. As each model is different, there is no single recording strategy that is ideal for all models. However, the concepts of activity, passive entity, interactions, and operations allow us to map all nondeterminism to a uniform trace format. Specifically, we propose a concurrency-model-agnostic event representation and recording mechanism. In the remainder of

---

[3] Non-concurrent programs can also be nondeterministic. We consider this nondeterminism as external input and sketch a strategy for handling it in section 7.

**Capturing Nondeterminism for Concurrency Model Agnostic Record & Replay**

■ **Table 1** Key concepts and sources of nondeterminism in each of the concurrency models.

|  | T&L | CEL | CSP | STM |
| --- | --- | --- | --- | --- |
| Activities | threads | actors | processes, | threads |
| Passive Entities | conditions, locks | promises | channels | - |
| Nondeterminism | lock sequence, timeout flags | message sequence on mailboxes and promises | channel read/write sequence | commit sequence |

this section, we detail how these model-agnostic events work, and how they can be recorded by considering the nondeterminism either at the sender or receiver-side.

### 3.1 Model-agnostic Event Infrastructure

Our model-agnostic record & replay framework is based on the idea that in a recorded execution each activity produces a sequence of trace events that represents the nondeterministic operations performed. This sequence of events reflects what happened in the original execution. During replay, each nondeterministic operation consumes trace events to reproduce the originally observed ordering and outcome.

In our approach, each nondeterministic operation is instrumented individually for record & replay. Furthermore, each trace event is produced and consumed by the same instrumented operation, which means each operation can define individually how to establish and reproduce the ordering. This offers the flexibility to select an appropriate strategy that exploits the properties of each nondeterministic operation in a concurrency model.

In multi-paradigm concurrent programs, any nondeterministic operation could be performed and recorded by any activity, e.g., a thread can send a message to an actor. These cross-model interactions are supported by making the operations responsible for the record & replay semantics and by using a uniform trace format.

**Event Format**   Each event consists of an event type and data that can be used to reproduce its ordering. The event type enables the instrumentation of an operation to capture additional information. For instance, the event type can encode whether an await on a condition variable timed out or returned normally. Our prototype uses a 1-byte event type that allows us to distinguish up to 255 events, followed by 8-byte of event specific data. The chosen sizes are suitable to support the concurrency models in this paper, but can of course be adapted if needed.

**Record & Replay Framework**   Our record & replay framework provides the basic functionality and builds on the uniform trace format for the recording and retrieving of events. A nondeterministic operation uses the recordInteraction() method (see listing 1), to record event type and ordering information. Buffer management and serialization are handled by the framework. For the parsing of a trace during replay, no knowledge about the concrete concurrency models is needed. Each activity accesses its trace

■ **Listing 1** Pseudo code for the key elements of our record & replay approach.

```
1  void recordInteraction(eventType, data) {
2    if (MODE == RECORD) {
3      currentActivity.buffer.putByteAndLong(
4        eventType, data); } }
5
6  void incrementVersion(entity) {
7    if (MODE == RECORD || mode == REPLAY) {
8      entity.version++; } }
9
10 void delayInteraction(entity, eventType) {
11   if (MODE == REPLAY) {
12     event = currentActivity.getNextEvent();
13     assert event.type == eventType;
14     while (entity.version != event.version) {
15       delay; } } }
```

■ **Listing 2** Pseudo code of instrumenting a lock for record & replay.

```
1  public class Lock implements Entity {
2    long version;
3
4    public void lock() {
5      delayInteraction(this, LOCK);
6      //implementation lock held afterwards
7      reentrantLock.lock();
8      recordInteraction(LOCK, this.version);
9      incrementVersion(this); } }
```

events as a queue with poll() and peek() operations. When an operation is replayed, it retrieves the next event from the queue and use the event's ordering information to reproduce the order of the original execution.

For our prototype we integrate the shared infrastructure and the concrete instrumentations of nondeterministic operations required for record & replay as an optional feature directly in SOMNS. This allows for transparent record & replay without need for program transformation.

### 3.2 Capturing Nondeterminism

For nondeterministic operations, we can establish and record an ordering of events with one of the following approaches: we can either capture the event when an activity initiates it (*sender side*) or when an activity or passive entity receives it (*receiver side*).[4]

For *sender-side recording*, the activity initiating the interaction records the event. As a result, the recorded interactions with a specific entity are scattered over multiple activities. To establish an ordering between them, we record a version number for the event. The management of version numbers can often reuse existing synchronization in the instrumented operations. The use of thread-local buffers comes naturally for sender-side recording because events are captured exclusively by activities.

If sender-side recording with version numbers is used by multiple models, it increases the uniformity of record & replay, and common operations (see listing 1) can be part of the framework.

In *receiver-side recording*, we record events in the entity that is interacted with. Here we can establish the event order by capturing the identity of the activity that causes the event. For instance, a lock records the sequence of threads that acquire it. Since passive entities can record events, but are not directly associated with a thread and its thread-local buffer, recording events is more challenging. Especially if the number of passive entities is high, it is inefficient for them to have their own buffers.

---

[4] For some models, for instance STM, the notion of sender/receiver is unusual. Here the active thread is the sender, and the global commit lock is the receiver.

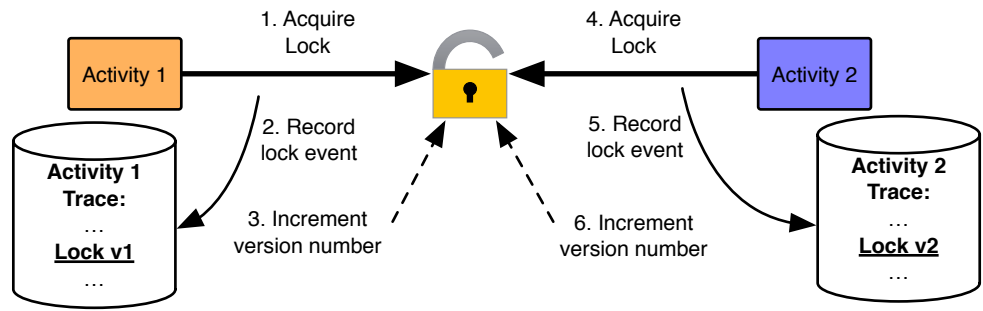

**Figure 2** Illustration of recording a program execution with two activities competing for a lock. Activity 1 acquires the lock first, and thus accesses the lock at version 1. The events are recorded into the corresponding traces of the activities.

To discuss the tradeoffs, we implement both receiver-side and sender-side record & replay for CEL actors in section 4.2. For the other concurrency models, we use sender-side recording as the implementation with thread-local buffers is straightforward.

## 4 Applying Concurrency-Model-Agnostic Record & Replay to four Concurrency Models

In this section, we discuss the implementation of our concurrency-model-agnostic record & replay approach for the concurrency models selected in section 2.3.

### 4.1 Record & Replay of Locks

Our record & replay for threads & locks uses sender-side recording. As outlined in section 2.3.1, one possible approach is to reproduce the order in which threads acquire a lock. Listing 2 shows the corresponding instrumentation of our lock implementation. SOMns wraps Java's ReentrantLock to do the actual locking (line 7). After lock() returns, the lock is held synchronizing the access to the version number. We then record the version number (line 8) and safely increment it (line 9). Figure 2 shows the lock acquisitions of two competing activities in a recorded execution.

The replay has to ensure activities acquire the locks in the recorded order. To this end, we delay the lock() call until the version matches using delayInteraction on line 5. The delay happens before the synchronization. Otherwise, a version mismatch would mean a thread holds the lock indefinitely causing a deadlock. Figure 3 depicts an example, where the first activity has to wait for another one before acquiring the lock.

**Condition Variables** In the presence of condition variables, implicit lock acquisition needs to be recorded and replayed as well. Since the lock associated with a condition needs to be held to perform await() or signal(), replay of lock acquisitions implicitly reproduces the order in which activities perform await() and signal() on a condition variable. Listing 3 shows an instrumented condition variable and its await() method. The return from await marks an implicit lock acquisition after being signaled. For record & replay we have to ensure that the interleaving of explicit and implicit lock acquisitions is the same. We do this by incrementing the version of the lock after await returns. The version increment prevents explicit lock acquisitions from acquiring the lock too early, i. e. they may have to wait for implicit acquisitions.

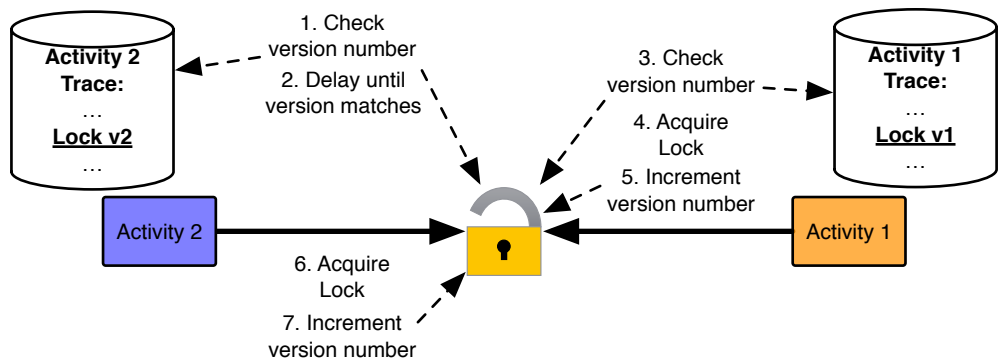

**Figure 3** Illustration of replaying the previously recorded execution. This time, Activity 2 started, but to preserve the recorded order, the acquisition of the lock is delayed, and Activity 1 can acquire it first as specified in the trace.

Record & replay of await with a timeout (awaitTimeout() in line 10) also needs to reproduce its outcome. To this end, we record different events (line 23) indicating whether the await timed out (AWAIT_TIMEOUT) or returned normally (AWAIT_SIGNALED). In both cases the lock is reacquired, and its current version is recorded as part of the event. Afterwards, the lock version is incremented (line 26). During replay, we simulate the outcome of awaitTimeout(). We first look at the next event in the activity's queue (line 13), if it is an AWAIT_SIGNALED event, we perform a regular await() without timeout (line 15). Otherwise, we simulate the timeout by releasing the lock and waiting for the right lock version to wake up (line 19). We do this by adding a dedicated condition variable which is used to wait until the version matches the recorded timeout version. Whenever the lock version is incremented, a signalAll() is performed on the replay condition (line 8) to wake up all waiting activities so they can check if they can continue.

### 4.2 Record & Replay of Communicating Event Loops (Actors)

In this section, we show both sender-side and receiver-side recording for CEL actors. Section 7 will discuss their tradeoffs including performance aspects.

**Sender Side**  Our sender-side record & replay handles the actor model's nondeterminism by treating the delivery of messages to an actor similar to lock acquisitions. Each actor has a version number to establish an ordering between the send operations. Listing 4 shows that the sender records an event (line 12) with the receiver's version number for each message. Afterwards, the version number is incremented (line 15).

In replay executions we need to ensure that messages are processed in the order indicated by their send event.  Since SOMNS uses a thread pool rather than assigning a dedicated thread to each actor, there is only a limited number of threads available for

**Capturing Nondeterminism for Concurrency Model Agnostic Record & Replay**

■ **Listing 3** Pseudo code implementing record & replay for awaiting a condition variable.

```
1  public class Condition {
2    Lock lock;
3  
4    public void await() {
5      doAwait();
6      incrementVersion(lock);
7      if (MODE == REPLAY) {
8        lock.replayCondition.signalAll(); } }
9  
10   public boolean awaitTimeout(int ms) {
11     boolean result = false;
12     if (MODE == REPLAY) {
13       event = currentActivity.getNextEvent();
14       if (event.type == AWAIT_SIGNALED) {
15         doAwait(); //no timeout
16         result == true;
17       } else { //simulate timeout: release lock,
18              //await version change
19         while (lock.version != event.version) {
20           lock.replayCondition.await(); } }
21     } else {
22       result = doAwait(ms);
23       recordInteraction(
24         result ? AWAIT_SIGNALED
25             : AWAIT_TIMEOUT, lock.version); }
26     incrementVersion(lock);
27     if (MODE == REPLAY) {
28       lock.replayCondition.signalAll(); }
29     return result; } }
```

■ **Listing 4** Pseudo code for actor message sending for record & replay.

```
1  public class Actor {
2    int version;
3    LinkedList<Message> mailbox;
4  
5    void send(message) {
6      synchronized(mailbox) {
7      if (MODE == REPLAY) {
8        event = currentActivity.getNextEvent();
9        assert event.type == MSG_SEND;
10       //store ordering in message
11       message.version = event.version; }
12     recordInteraction(MSG_SEND, version);
13     mailbox.add(message);
14     if (MODE == TRACING) {
15       incrementVersion(this); } } } }
```

the actor program to make progress. Hence, using delayInteraction to delay a message send to the right time could cause deadlocks.

Instead of delaying send operations, we attach the version number of the message send to the messages (line 10), and order the mailbox with a priority queue. The message at the head of the queue is removed and processed only when no messages with a smaller version can still arrive. Using a counter to generate version numbers means the version is identical to the number of messages before a message.

As discussed in section 2.3.1, there is a race between sending messages to and resolving of a promise. Depending on the promise resolution state, either the resolver of the promise or the original sender of the message insert it into the mailbox. Therefore, the send event could be recorded by either activity. To solve the ambiguity, we record events and their order for operations on promises, i. e. storing a message and resolving a promise. These events ensure that promise messages are delivered to the mailbox of the receiver with the right version attached.

**Receiver Side**  For receiver-side record and replay of actors we adapted our previous work [3] to the concurrency-model-agnostic record & replay. The changes were limited to switching to the uniform trace format and the model-agnostic infrastructure. Pseudo code illustrating the changes is shown in listing 5 in the appendix.

Our receiver-side record & replay distinguishes between "normal" messages that are sent directly from one actor to another, and *promise messages* that are sent to a promise (unresolved or resolved). Before an actor starts processing a message in a traced execution, we record an event that contains the ID of the actor that sent the message (line 9). To capture their ordering, promise messages are sequentially numbered by their sender, combined with the ID of the sender this allows us to uniquely identify them. The sequence number is recorded in a separate event (line 7) just before the message send event we record for all messages. Hence, the data for ordering promise messages is split into two separate events. By recording the promise message event first we are able to determine what kind of message was received and accordingly use one or two events during replay.

In replay executions the `processMessages` method (line 24–35) reorders messages. The first message that matches the next event(s) in the execution trace (line 29) is removed from the mailbox and processed (line 31).

### 4.3 Record & Replay of Communicating Sequential Processes

Record & replay for CSP can be achieved by reproducing the total order of reads and writes for each channel that is subject to races (see section 2.3.1). However, as our CSP implementation uses a rendezvous semantics, a total order is not necessary and a partial order of events that reflects the order of rendezvous is sufficient. We attach a version number to each channel. For all reads and writes performed on a channel an event with the version number is recorded. The version number is incremented only after a rendezvous is completed. In replay executions, read and write operations are delayed until the version number of the recorded event matches the channel. A more detailed description of instrumenting CSP is available in appendix A.2.

### 4.4 Record & Replay of Software Transactional Memory

As discussed in section 2.3.1, we consider a side-effect free STM, i. e. each transaction is retried until it succeeds and retries are not observable by the program. Therefore, nondeterminism manifests only in the order in which transaction are committed. Record & replay of such STM systems requires only to recreate the recorded commit order. We use a global commit version number, which is recorded and incremented whenever a transaction is committed successfully (see listing 7 in the appendix).

During replay, when an activity starts the commit, it first performs the conflict check. If there are no conflicts, we check if this activity is the next that is supposed to commit something according to the trace. To this end, we check the event at the head of the activity's event queue, and compare the version stored in it with the global commit version. If the two numbers are equal, the activity can remove the event from the queue and commit the transaction's changes to shared memory. If the versions do not match, the transaction fails, and starts again from the beginning. This is repeated until both the conflict check passes and the global commit version matches.

Note that this replay does not reproduce the number of failed attempts the original execution had, nor does it reproduce what happens inside failed transactions. This is an acceptable strategy for STM implementations that retry transactions indefinitely (as in SOMns). For STM systems in which failed transactions are exposed to the program, the record & replay needs to be designed differently (see section 5). In either case, our approach is agnostic to the details and provides support for recording & replaying of the necessary events.

## 5 Evaluation of Expressiveness and Flexibility

This section evaluates the flexibility and generality of our approach. We aim to support a wide range of concurrency models, vary the record & replay strategy, and how to model events without having to change the trace format or its infrastructure.

**Expressiveness of Record & Replay** Our prototype supports deterministic replay for all four concurrency models we set out to support in section 3.1. Section 2.3.1 identified the underlying nondeterminism (see table 1), and section 4 devised a concrete implementation for each model, demonstrating for instance that sender and receiver-side recording are both supported by our approach.

Implementing record & replay presents different challenges for each model, in the simplest case the nondeterministic events of the model are simply mapped to recorded events. However, for concurrency models such as the CEL model, identifying a suitable mapping can be more involved. Depending on whether a sender or receiver-side recording is chosen, we needed to devise a strategy for instance to avoid blocking actors in replay, which otherwise could lead to deadlocks. As demonstrated, a non-blocking solution was possible using the uniform trace format, even without changing the event recording. We argue that this demonstrates that our concurrency-model-agnostic record & replay is expressive since there are multiple ways to record & replay event orders with a common infrastructure and trace format.

**Flexibility of the Trace Format Encoding** Our trace format consists of a 1-byte header followed by 8 bytes of event data. As outlined in section 3.2, we typically use the 8 bytes to store long version numbers. However, these 8 bytes can be used arbitrarily for each type of event, as it is interpreted and used locally when trace events are consumed during replay. For instance, the 8 bytes can store a unique ID of an entity, a boolean result, or two 32-bit floating point numbers. If some type of event needs more than the available 8 bytes, for instance in our receiver-side implementation for CEL actors (see section 4.2), consecutive events can be recorded and the larger data split between them. However, since our trace format uses a uniform data representation for all events, the parser is kept minimal and independent of specific event formats.

Arguably, this approach strikes a good balance, making the trace format uniform and compact (see section 6.2), while also providing the flexibility to encode data for events as needed. With this separation, only the code handling a specific event needs to know how to record or interpret the data of the event for replay.

**Conclusion** Given the successful implementation of the concurrency models we set out to support in section 2.2, and the flexibility for different mapping strategies, we argue that our record & replay is able to support a wide range of models and their variations. This is because the identified and recorded events are a foundation for supporting other variations of the four models, too. Furthermore, we argue that supporting a new model, from the multitude of special-purpose concurrency models, is a process of identifying relevant nondeterministic events, mapping them to an event encoding, and then devising a decoding strategy. For example, record & replay of STM systems with limited retries or weaker isolation may have to consider the order of read/write operations in addition to the order of commits. Though such variations are supported by the infrastructure for record & replay, which can be reused from the existing concurrency models.

# 6 Evaluation of Performance

The goal of our performance evaluation is to assess whether our approach is practical, and how its compares to a special-purpose implementation.

To assess practicality, we measure the run-time overhead of recording for each model individually as well as for our multi-paradigm application. For a diagnostic use in production scenarios, we assume an overhead of 10 % may still be acceptable [3], while doubling the run time may disqualify the approach from consideration. To demonstrate that the system indeed supports multi-paradigm applications, we also measure the recording overhead for our application of section 2.1.

To compare our approach to a special-purpose implementation, we use our previous work on record & replay for actors [3]. It optimized recording to achieve minimal run-time overhead for a real-world application. As such, we consider it to be a well optimized baseline for comparison, which is also built on SOMns.

## 6.1 Methodology

We evaluate the recording performance of our implementation in SOMns based on different benchmark suites for each model. For the CEL model, we rely on the widely used Savina benchmark suite for actors [16]. For threads & locks and STM we use the LeeTM benchmark [1], the Vacation benchmark from the STAMP suite [26], and a variant of the classic dining philosophers. Lee and Vacation were designed for transactional systems. For our threads & locks version, we replaced transactions with mutual exclusion through locks. For CSP, we adapted a few of the Savina benchmarks.

To demonstrate the support of multi-paradigm applications, we implemented the sales processing system discussed in section 2.1.

For the CEL model, SOMns comes with a comprehensive set of benchmarks enabling a detailed evaluation. However, for the other concurrency models, we had to port benchmarks, which limits the evaluation. Furthermore, SOMns' STM implementation is not optimized. This means its evaluation is merely an indication that our approach works, but does not generalize to the performance overhead for an optimized system.

**Capturing Nondeterminism for Concurrency Model Agnostic Record & Replay**

**Benchmark Execution**  SOMns is a suitable platform for the evaluation of run-time overhead, because it has performance comparable to the widely used NodeJS [23]. SOMns achieves this performance by using the Graal just-in-time compiler [36]. Because of this run-time profiling and compilation, benchmarks do need time to *warm up* [7], which depending on the benchmark, may take multiple iterations before performance stabilizes. We account for this warmup behavior by executing each benchmark for 2000 iterations. After manual inspection, we discarded the first 500 iterations as warmup. The remaining 1500 iterations of each benchmark are assumed to be representative for the performance behavior of a longer running application.

The overall overhead of recording traces for a program consists of two components: the instrumentation that produces the trace data, and the writing of traces to disk. A SSD may be able to write the traces produced by a benchmark with close to zero overhead, while a HDD with much lower writing speed may represent a bottleneck that significantly increases overhead. For the Savina benchmarks, using a HDD instead of a SSD on average results in a 30.52 % overhead (min. −0.23 %, max. 107.67 %). To avoid conflating disk speed and the intrinsic overhead of our approach, we measure only the overhead caused by the tracing itself. We do the same for the specialized actor record & replay to keep results comparable. This means, buffers are written to during tracing, but the background thread that would write them to disk, instead resets them. Results that include writing traces to a HDD and a SSD are in the appendix in figure 13.

We generally report the *run-time factor* scaled based on the performance of the baseline without recording support as it allows us to compare the results intuitively.[5]

The benchmarks were executed on an octa-core AMD Ryzen 7 3700x CPU, 3.60 GHz, with 32 GB RAM, a 256 GB SSD, Fedora 32 (kernel 5.7.15-200), and a custom built OpenJDK 1.8.0_232 with JVMCI and Graal version pre-20.3. We used the ReBench tool [22] to run the benchmarks and collect the results. ReBench reduces noise in benchmark results by disabling frequency scaling, turbo boost, and similar CPU features. Furthermore, ReBench reserves CPU cores and supports thread pinning to avoid threads jumping between different CPU cores, for instance as a result of GC.

In a multithreaded execution, the performance of a benchmark may be sensitive to contention. For instance in the Philosophers benchmark, activities compete for access to forks, depending on the order in which philosophers acquire and release forks the number of failed attempts and the time the benchmark takes may increase. As our tracing of nondeterministic events inevitably is going to have an effect on the execution, we restrict execution to one thread to obtain more stable results that reflect the actual tracing overhead. The single threaded execution ensures that the tracing overhead is not hidden, e.g., by thread contention. However, for the CSP benchmarks, we had to use at least two threads for execution, since the processes are directly mapped to threads and channels have rendezvous semantics. The CSP benchmarks do not have contention on the channels when only two threads are used. Therefore, any overhead of tracing should still be measurable. Similar to the CSP benchmarks, we did not restrict parallelism for the multi-paradigm sales processing app. For completeness, we include multithreaded benchmark results in the appendix in figure 14.

---

[5] The run-time factor is a simple scaling transformation, which does not affect statistic properties of the data, similar to a transformation from °C to °F. It scales each measurement to the mean measurement for the benchmark on SOMns without recording support.

## 6.2 Recording Performance for CEL Compared to Special-Purpose Solution

The recording performance of the CEL model is evaluated with the Savina actor benchmarks. We assess runtime overhead and trace size of our concurrency-model-agnostic record & replay with the specialized CEL record & replay system [3] that is optimized for performance. Since both approaches are implemented on SOMns, we can directly compare our approach (Sender-Side) with this special-purpose solution (Specialized). Additionally, we compare to the reimplementation of the specialized recording (Receiver-Side) on top of our concurrency-model-agnostic approach.

**Run-time Performance**   Figure 4 shows the recording overhead, i. e. the run-time factor, of the three implementations normalized to untraced execution. Figure 5 gives a summary of the overall performance for each of the record & replay variants.

For recording nondeterministic events of the CEL model with our concurrency-model-agnostic approach (Sender-side) the geometric mean of the run-time factors (red dots in the plot) indicates an overall overhead of 9.87 %. ForkJoinActorCreation has the highest overhead with 22.93 %. In contrast, the ForkJoinThroughput benchmark has an overhead of 0.03 % run-time compared to SOMns without recording support.

The benchmarks indicate performance trade offs between the approaches. For instance, sender-side recording is faster than the specialized solution for the CobwebbedTree benchmark, but slower for BankTransaction. The specialized actor record & replay has an average overhead of 8.9 % (min. −0.69 %, max. 26.18 %), while our sender-side approach had an overhead of 9.87 % (min. 0.03 %, max. 22.93 %). This indicates that the recording overhead of both approaches is in a similar range and that our approach is competitive with an optimized special-purpose implementation.

With an average of 13.18 % (min. −0.91 %, max. 34.34 %) the recording overhead of our receiver-side implementation is higher than the specialized one (avg. 8.9 %), as we do not have the same optimizations.

Our results indicate that the recording overhead depends on the mix of operations performed by an application. Benchmarks with a high proportion of nondeterministic events such as ForkJoinActorCreation, which just creates actors and sends messages to them, have higher overhead than benchmarks with few nondeterministic events such as Trapezoidal and NQueens. For instance, Trapezoidal is one of the more computationally intensive benchmarks of the Savina suite, and thus, has a comparably small number of nondeterministic events that are recorded.

Figure 9 in the appendix details the performance of the untraced baseline. The multithreaded performance is also available in the appendix as figure 14.

**Trace Size**   Figure 6 compares trace sizes. A detailed table of trace sizes is in appendix table 2. The trace sizes of our receiver-side implementation are close to those of the specialized actor record & replay approach, being on average 3 % larger (min. 0 %, max. 6 %). This is expected and the difference is caused by the additional event headers for the promise message events. In contrast, the sender-side approach produces traces that are on average 24 % larger (min. −34 %, max. 111 %). The results show that there are tradeoffs between the recording strategies, and the best choice depends on the type of nondeterminism exhibited by the benchmark.

We conclude that our record & replay can reach performance and trace sizes competitive with approaches optimized for a single model, and is thus practical.

### 6.3 Recording Performance for Threads & Locks

The recording overhead on the T&L model is measured with benchmarks designed to compare locking with STM, namely the LeeTM and Vacation benchmarks, both about 600–700 LOC each. We also ported the Philosophers benchmark. The benchmarks were intended to measure how the nondeterministic program parts are synchronized, and how the synchronization overhead changes performance. Since we run the benchmarks on a single thread, the synchronization overhead is not relevant. However, the nondeterministic events originate from the same program elements, and thus, allow us to measure their overhead.

Figure 7 shows the overall recording performance for our benchmarks of threads & locks. The geometric mean of the run-time factors is 7.85 % (min. −0.31 %, max. 17.84 %). This means, the results are in a similar range of overhead as the record & replay for the CEL model. For individual benchmark results we refer to figure 10 in the appendix.

### 6.4 Recording Performance for CSP

We evaluate the recording overhead for the CSP model based on adapted Savina benchmarks. While programs designed for CSP may take different shapes, the benchmarks show a variety of message passing behaviors that can also appear in CSP programs.

As shown in figure 7, recording executions for our CSP benchmarks changes the run time on average by 21.82 % (min. −12.5 %, max. 87.15 %). Despite our best efforts to minimize external impacts on the benchmark execution, the results are very noisy, especially for the ForkJoinThroughput benchmark. For more detailed results we refer to the appendix (figure 11)

### 6.5 Recording Performance for STM

The evaluation of the recording overhead on the STM model uses the same benchmarks as threads & locks. The overall benchmark results are shown in figure 7 and detailed results are available in figure 12 in the appendix. The geometric mean of the run-time factors indicates a run time change of 0.39 % (min. −0.14 %, max. 0.94 %). These numbers are unfortunately not generalizable, as the STM implementation of SOMns is not optimized for performance. While the other concurrency models execute optimized code, the STM is currently about 5× slower than sequential code. This means, the STM itself hides any overhead from the recording. However, our approach to recording for the STM records only one event per successfully committed transaction. Thus, if we had applied it to a highly optimized STM, it would have a small constant overhead for each commit operation, which would likely be in the low percentage range for applications with large number of transactions.

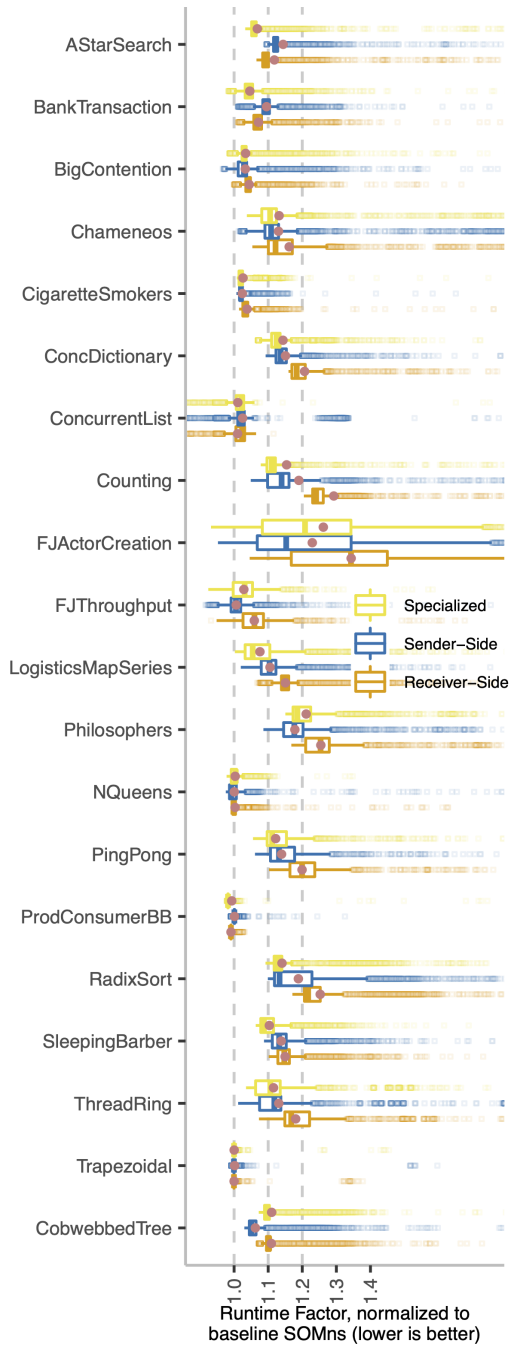

**Figure 4** Recording performance of our record & replay for CEL actors compared to regular benchmark executions. The average overhead is 9.87 % (min. 0.03 %, max. 22.93 %) and thus, in an acceptable range for many production applications.

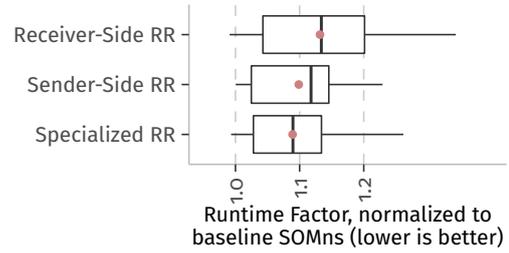

**Figure 5** Summary of the recording performance of our record & replay for CEL actors compared to regular benchmark executions.

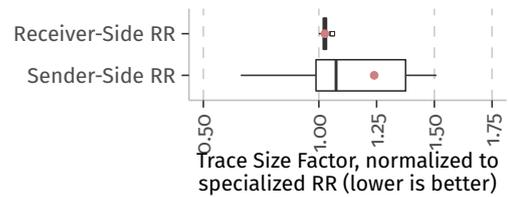

**Figure 6** Trace sizes of the Savina benchmarks for both sender- and receiver-side record & replay, normalized to the specialized record & replay.

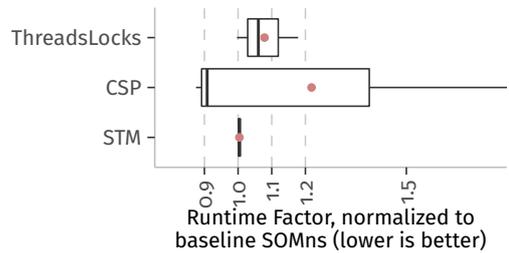

**Figure 7** Recording performance of our prototype for various models compared to untraced execution.

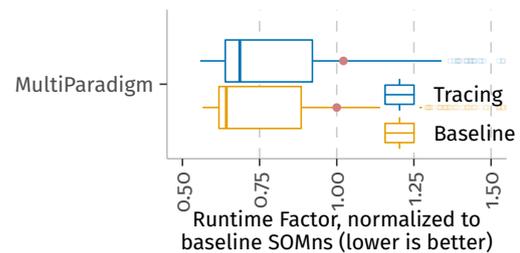

**Figure 8** Recording performance of our prototype for a multi-paradigm workload compared to untraced execution.

### 6.6 Recording Performance for a Multi-Paradigm Program

As described in section 2.1, the sales processing system combines concurrency models by having each of the three subsystems rely on different concurrency model. The benchmark uses four actors: one to simulate JSON input events, and the other three representing the JSON parsing, storage, and forecast subsystems. The JSON parsing itself uses two CSP processes communicating with channels. The storage system is integrating with a main system, but in this benchmark simply uses a STM transaction to update data structures with the incoming sales data. Once all data is received, the forecast actor receives a message with the data and spawns threads to calculate a linear regression to do a simple sales trend estimate. The benchmark completes an iteration when it receives the forecasts and confirms that they are the expected values.

The recording performance for our multi-paradigm benchmark, in figure 8 indicates that the average arithmetic run time changes by 2 %. Unfortunately, the noise prevents us from drawing conclusions, but it demonstrates that we support such applications.

## 7 Discussion and Future Work

This section discusses design decisions and alternatives. It also identifies future work.

**Trade-off Between Sender-side and Receiver-side Recording**  In section 4.2, we presented two alternative strategies to record & replay the nondeterminism in the CEL model. Besides the recording perspective, the implementation of the two strategies also differs in how promise nondeterminism is handled. For the receiver-side implementation, the nondeterminism is implicitly resolved and not an issue. The sender-side variant needs to handle promise nondeterminism explicitly, which significantly increases code complexity. The performance results in figure 4 show that the sender-side recording had a lower overall overhead, but we believe that the receiver-side recording offers more optimization opportunities. Hence, there is a trade-off between different instrumentation strategies in terms of code complexity, optimization opportunities, and performance. We recommend to weight the benefits and drawbacks when it comes to deciding on sender or receiver-side recording. Arguably, our concurrency-model-agnostic record & replay approach supports both, and thereby gives the flexibility to choose the preferable option.

**More Optimal Event Encoding**  Our chosen event encodings are not necessarily optimal. For instance, for threads & locks, consecutive lock acquisitions by the same thread could be captured more efficiently, and one could avoid recording events that do not influence other threads. As demonstrated by our two strategies for CEL, we support many different encodings and using techniques to further reduce recording overhead and trace size are subject of future work.

**Capturing External Nondeterminism** Nondeterminism in programs is not limited to timing and scheduling of concurrent events; input from the environment influences a system, too. For example, the incoming HTTP requests in a web application represent a source of nondeterminism as in different executions the number of clients, the type and content of requests varies. Similarly, applications that use databases or access files cannot be accurately replayed if during replay execution the files/database have a different state. Capturing such external inputs is essential for deterministic replay.

In previous work [3] we reproduced external nondeterminism in actor systems using an event-based system that tracks external influence on each actor. The system distinguishes system calls and asynchronous data sources as two sources for nondeterminism. For each system call, an event is recorded that contains a reference to the serialized result of the system call. Asynchronous data sources use messages sends, which are marked and recorded as external messages including their data for replay

This approach to deal with external inputs can be combined with the record & replay approach presented in this paper. System call events could be recorded as an event in our uniform trace format with a one byte header, followed by the ID of the serialized data. Recording of data from asynchronous sources can be done with a dedicated mechanism to serialize the nondeterministic inputs efficiently.

**Support for Shared Memory Accesses** In this work, we focused on high-level nondeterminism of the concurrency models. As a result, our current implementation relies on the assumption that all shared memory accesses are synchronized. Existing literature [4, 9, 19, 20, 21, 27] describes various strategies and optimizations to avoid record & replay of individual memory accesses and may provide a starting point for supporting them in our prototype. Typically, forced single processor execution [4, 27] or grouping of accesses are used to either completely remove races between individual memory accesses or at least coarsen the granularity of recorded events. iReplayer [21] follows a different strategy, they do not record unsynchronized shared memory accesses and deal with the nondeterminism by repeatedly replaying the program until the right execution was found. In the future, we may add a mechanism to automatically version and synchronize shared objects by tracking object sharing [10]. This would allow us to record only accesses to memory that is used by multiple activities.

**Correctness of the Approach** This work does not claim any formal notion of correctness. However, Torres Lopez, Gurdeep Singh, Marr, Gonzalez Boix, and Scholliers [32] investigated a formalism to model debugging, and applied it to debugging actor programs. They proved non-interference between a debugger and base language. Future work could use it as a starting point to prove the correctness of our record & replay approach.

Our replay implementation uses assertions that check that the types of events in the trace match the execution. The implementation is tested and runs the benchmarks and record & replay as part of the test suite on our CI system. This detects a certain class of bugs, including undesired nondeterminism. Incorrect replay quickly leads to deadlocks or assertion errors. Thus, while we do not prove correctness, we are confident that our implementation is correct enough to be practical.

**Generalizability**   In our implementation, we directly instrument the concurrency model implementations in the SOMns interpreter and leverage Truffle/Graal with its partial evaluation for performance. However, neither are requirements for the proposed record & replay approach. Instrumentation can also be achieved by other means, for instance for Java one may use bytecode transformation during class loading. While using bytecode transformation to support numerous concurrency libraries is more challenging, we believe that it can achieve similar results in terms of functionality. To reach the same performance, some form of cooperation with the compiler will be needed to clearly identify the operations that do not depend on run-time data and to optimize them as much as possible. The literature on shared memory record & replay already provides alternative strategies to this effect [4, 9, 19, 20, 27].

**Usefulness of the Approach**   SOMns is a research language implementation that is not used by any production systems. However, record & replay was useful for us for fixing bugs in the replay implementation itself. For instance, a trace containing a rare corner case allowed us to repeatedly debug that corner case and fix the problem, which corresponds to the typical use cases for such systems.

## 8   Related Work

While literature in debugging is extensive, concurrency model agnostic debugging support is rare. In particular, record & replay debugging has been explored for a variety of systems [3, 4, 9, 13, 19, 20, 27] with different goals and strategies. However, the existing tools are designed for individual concurrency models rather than multi-paradigm concurrent scenarios. For brevity, we focus our discussion on the most closely related work.

**Kómpos**   In prior work we have investigated online debugging of multi-paradigm concurrent programs for the Kómpos debugger protocol [24]. It allows language runtimes to define custom breakpoints and stepping operations for concurrency models without having to adapt a debugger. While both Kómpos and our approach use the same entity abstraction and the events, their representation and purpose differ. For online debugging, more information is needed than for an efficient record & replay, which only needs to reproduce the order of nondeterministic events.

**Instant Replay**   A classic approach to record & replay is Instant Replay [19]. In Instant Replay, a concurrent read exclusive write (CREW) protocol is used to record a partial order of accesses to shared objects. Each shared object has a version number and a semaphore. Read operations record the current version number of the object, as multiple reads can record the same version number, a counter tracks how many times a version was read. In Instant Replay, each thread records events on a *tape*. When a read or write is performed during replay, the next event, i. e. version number, is read from the tape. If the version number does not match the version number of the object that is to be accessed, execution blocks until the version number matches. In addition to the version number check, write operations have to ensure that all reads that depend on this version have been performed before the object can be written to.

Instead of object accesses, we record generic nondeterministic events with a uniform and flexible trace format that provides more information than version number tapes.

There is also a version of Instant Replay that uses Lamport clocks instead of version numbers to produce smaller traces [20]. Future work could examine the performance and trace size implications of using Lamport clocks instead of version numbers.

**Shared Memory Record & Replay** Tardis [4] and rr [27] provide record & replay debugging for shared memory applications. Tardis also supports time-travel debugging by integrating with the GC and recording snapshots at regular intervals. Both implementations handle shared memory races by executing all threads of an application on a single processor core. To achieve deterministic replay, they record system calls and asynchronous nondeterministic events, for instance context switches and signals.

An alternative approach to avoid recording shared memory access in a multi-threaded setting is described by iReplayer [21]. They accept that replayed executions are not entirely deterministic and attempt to replay the execution multiple times until the result of the shared memory races is identical.

In contrast to Tardis and rr, and similar to iReplayer, we focus on the high-level concurrency constructs. Furthermore, record and replay can use multiple cores. However, this currently excludes handling of shared memory accesses, which we leave to future work (see section 7).

**Event Loop Record & Replay** The authors of the Tardis [4] time travel debugger also applied their approach to JavaScript event loops to develop Jardis [5], which is based on a modified JavaScript engine. As JavaScript event loops are traditionally single threaded, nondeterminism is limited to system calls, inputs, and scheduling.

While Jardis and our approach have a modified language implementation in common, they significantly differ in the nondeterminism that needs to be recorded as our approach supports multiple activities and multi-threaded execution.

# 9 Conclusion

Nondeterminism in concurrent systems makes them hard to debug. Today's record & replay systems only aid debugging of applications that use a single concurrency model. However, modern applications combine various models and are not supported.

To address this issue, we presented a concurrency-model-agnostic record & replay approach that captures high-level nondeterministic events of concurrency models to reproduce their order deterministically when replaying an execution. Our uniform

**Capturing Nondeterminism for Concurrency Model Agnostic Record & Replay**

and flexible trace format can represent a wide range of different events, which may be needed to implement record & replay for a particular model.

The evaluation shows that this approach is flexible enough to capture the nondeterministic events of four common concurrency models: threads & locks, communicating event loops, communicating sequential processes and software transactional memory. It also supports different implementation strategies, as showed by implementing sender as well as receiver-based record & replay for communicating event loops. Using this flexibility, our implementation in SOMns is able to record & replay applications that mix and match these four concurrency models. We demonstrated this by building a sales processing application that combines all four models.

The evaluation of the recording performance focuses on the CEL model. We compare the performance of our concurrency-model-agnostic record & replay approach with that of an optimized record & replay system built specifically for the CEL model. On the Savina actor benchmarks, our approach shows an average trace recording overhead of 10 % (min. 0 %, max. 23 %). This is on the same level as what we reported for the optimized and special-purpose CEL record & replay [3]. Furthermore, comparing the receiver-side implementations, the trace size of our approach is only 3 % (min. 0 %, max. 6 %) larger on average than that of the specialized one.

For the other concurrency models, we ported various benchmarks and found that the overhead is in a similar range. To demonstrate the support of recording multi-paradigm concurrent programs, we used the sales processing application as benchmark.

From these results, we conclude that our approach is general and flexible enough to accommodate a wide range of concurrency models with different types of nondeterministic events, and that its performance is comparable to a system optimized for a specific concurrency model. This means, our concurrency-model-agnostic approach enables us to debug modern applications more deterministically. This will simplify debugging and maintenance of applications that mix and match concurrency models.

**Acknowledgements** This research is funded in part by a collaboration grant of the Austrian Science Fund (FWF) and the Research Foundation Flanders (FWO Belgium) as project I2491-N31 and G004816N.

## A  Appendix of Code Examples for the Instrumentation for Record & Replay

In this appendix, we include additional listings of pseudo code to complement the description in section 4.

### A.1  Recording of Actor Message Receiving

In listing 5, we see the sketch of how receiver-side recording would instrument the processing of messages in an actor. As outlined in section 2.3.1, we need to treat the promises specially to capture all of their nondeterminism, which is done in line 7. For all messages, including promise messages, we record the sender ID in line 9.

The replayCanProgress() method (line 13) implements the checks to match the event, which is next in the trace to the received message. Only if it oof the expected type, and the sender matches, or in the case of a promise message also the message ID, replay will continue. This approach is possible, because a single sender has a deterministic order of messages it is sending.

**Listing 5** Pseudo code for instrumenting actor message receiving for record & replay.

```
 1  public class Actor {
 2    LinkedList<Message> mailbox;
 3    private int promiseMessageCnt;
 4
 5    void process(message) {
 6      if (message instanceof PromiseMessage) {
 7        recordInteraction(PROMMSG_RCVD, message.messageId);
 8      }
 9      recordInteraction(MSG_RCVD, message.sender.getId());
10      message.execute();
11    }
12
13    boolean replayCanProcess(message) {
14      event = currentActivity.peekNextEvent();
15      assert event.type == MSG_RCVD || event.type == PMSG_RCVD;
16      if (event.type == PMSG_RCVD) {
17        if (!(message instanceof PromiseMessage)) { return false; }
18        if (message.messageId != event.data) { return false; }
19        event = peekNextNextEvent(); //peek one further
20      }
21      return event.data == message.sender;
22    }
23
24    void processMessages() {
25      if (MODE == REPLAY) {
26        progress = true;
27        while (progress) {
28          progress = false;
29          for (message : mailbox) {
30            if (replayCanProcess(message)) {
31              mailbox.remove(message); process(message);
32              progress = true; break; }}}
33      } else {
34        while (!mailbox.isEmpty()) { process(mailbox.get()); }}
35    }
36  }
```

### A.2 Instrumenting CSP for Record & Replay

Listing 6 shows the instrumented channel read and write methods. When an activity performs a read or write operation on a channel, we record a corresponding event with the current version number of the channel (lines 7 and 14). The version number is incremented only once per rendezvous, e.g. after the write returns from the rendezvous (line 9). This means both the read and write events have the same version number. In a replay execution, this allows the reader and writer to arrive in arbitrary order for each rendezvous while ensuring a correct replay order. Read and write events are delayed until the version number present in the event queue of the performing activities matches the version number of the channel (lines 5 and 12).

■ **Listing 6** Pseudo code for instrumenting a CSP channel write for record & replay.

```
1  public class Channel {
2    int version;
3
4    void channelWrite(message) {
5      delayInteraction (this, CHANNEL_WRITE);
6      synchronized(this) {
7        recordInteraction(this, CHANNEL_WRITE);
8        doWrite(message); //blocks until rendezvous
9        incrementVersion(channel); } }
10
11   Object channelRead() {
12     delayInteraction(this, CHANNEL_READ);
13     synchronized(this) {
14       recordInteraction(this, CHANNEL_READ);
15       return doRead(message); } } //blocks until rendezvous
```

■ **Listing 7** Pseudo code for instrumenting STM commits for record & replay.

```
1  public class Transactions {
2    int version;
3    Object commitLock;
4
5    boolean commit() {
6      synchronized(commitLock) {
7        if (hasConflict()) { return false; }
8
9        //retry transaction during replay when version does not match
10       if (MODE == REPLAY) {
11         if (currentActivity.peekNextEvent().version != version) {
12           return false;
13         } else {
14           currentActivity.getNextEvent(); } } // consume event
15
16       recordInteraction(this, TRANSACTION_COMMIT);
17       incrementVersion(channel);
18       applyChanges();
19       return true; } } }
```

### A.3 Instrumenting STM for Record & Replay

Listing 7 sketches our instrumentation for SOMns's software transactional memory. Line 16 does the recording of the transaction during normal execution. In line 12, we see the case that the transaction was completed too early during reply. We simply consider this as a conflict having occurred, and retry the transaction.

## B  Appendix to Performance Evaluation

This section gives further details on the performance evaluation, covering experiments previously mentioned, but relegated here for brevity.

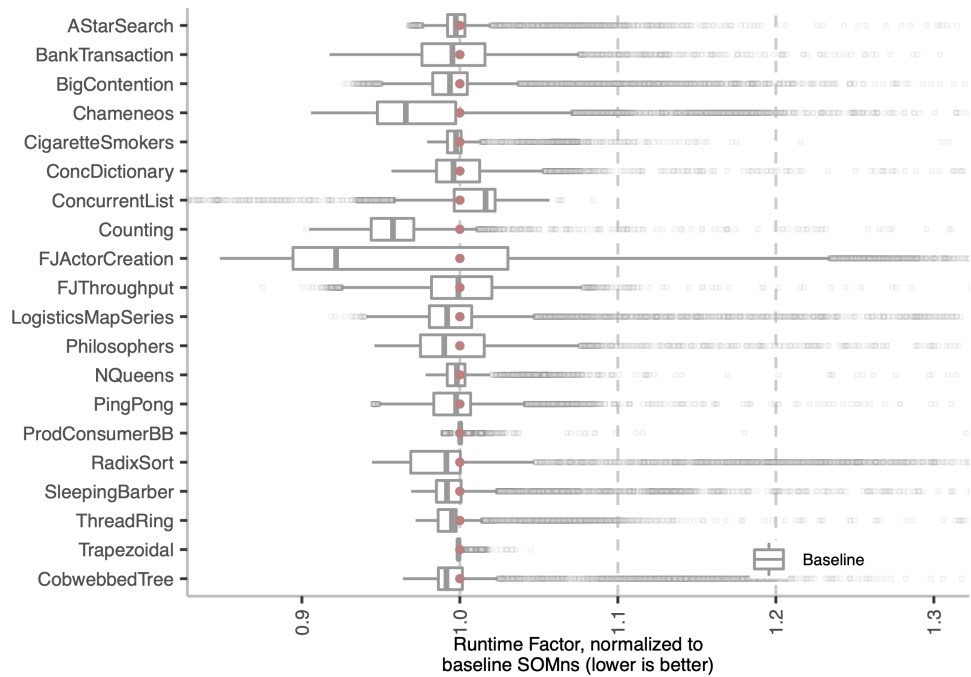

**Figure 9** Baseline performance of CEL actors for the Savina benchmark suite. The red dot indicates the geometric mean, to which all results are of a benchmark are normalized.

### B.1 Communicating Event Loop: Baseline Results

Figure 9 shows the behavior of the untraced baseline of CEL actors on top of SOMns. For each benchmark, all measurements are normalized to the geometric mean, which is indicated as red dot on the plot. What becomes visible in this chart is the differences in result distributions, and the spread of the normal behavior of the benchmarks. Benchmarks like the FJActorCreation, have a large range of behavior caused by the dependency on scheduling of the actors. Trapezoidal however, shows only minimal variation, because the benchmark has a much more deterministic structure.

### B.2 Threads & Locks: Results per Benchmark

For brevity, section 6.3 showed only a summary of the results for the recording performance of threads & locks. Figure 10 shows the results for all three benchmarks separately. We can see that the larger benchmarks Lee and Vacation are seeing a lower impact by tracing than the Philosophers microbenchmark. Generally, the impact is determined by the lock usage patterns, aspects such as contention and further instruction mix.

**B.3 CSP: Results per Benchmark**

Section 6.4 discussed the results based on an overview plot. Figure 11 shows the trace recoding overhead for the communicating sequential processes benchmarks in separation. Most notably, these benchmarks observe a lot of variation in their behavior, which is due to the nondeterministic nature of for instance the Philosopher benchmark. However, the tracing does neither affect their behavior much nor their performance. In larger applications, we expect the overhead to be a function of how frequently channel sends/reads are used compared to other operations. Compared to the results for threads & locks (figure 10), we see that the performance of microbenchmarks is also depending on the type of synchronization primitive. Since channel operations are slightly more elaborate in general, the overhead is here lower for Philosophers than when using a simple lock.

**B.4 STM: Results per Benchmark**

As discussed in section 2.3.1, SOMNS's STM system has a few specific properties, for instance that it does not abort transactions, but instead retries indefinitely. Furthermore, the STM system itself is not optimized, and thus, a bit of additional overhead introduced by tracing will not be visible.

For the performance of our trace recording, this means figure 12 indeed shows no significant overhead. However, because of the design choices, and missing optimizations, these results are not generalizable to other STM systems.

**B.5 Size of Traces Recorded**

Table 2 shows the size of the traces recorded for the Savina benchmarks. We compare the sizes of sender and receiver-side recording in our uniform trace format with the size of the record & replay system specialized for actor systems [3].

The table shows that the size factor, i.e., how much larger the sender or receiver-side encoding is than the specialized one, depends on the type of benchmark and the approach. For example, the trace for the Counting benchmark has only 2/3 the size when using sender-side recording than when using the specialized system. However, the receiver-side recording has a 6 % overhead. On the other hand, AStarSearch shows a 2.11× overhead with sender-side recording while only having 3 % overhead for receiver-side recording.

The Trapezoidal benchmark is also interesting, because it does not require any tracing in the receiver-side recording, since it does not have any nondeterminism from this perspective.

The specialized system uses an approach very close to the described receiver-side recording, and thus, the trace sizes are very similar. This shows that the uniform trace format has only a minimal overhead, but at the same time, gives the flexibility to also support sender-side recording, which can lead to much more compact traces, depending on the application details.

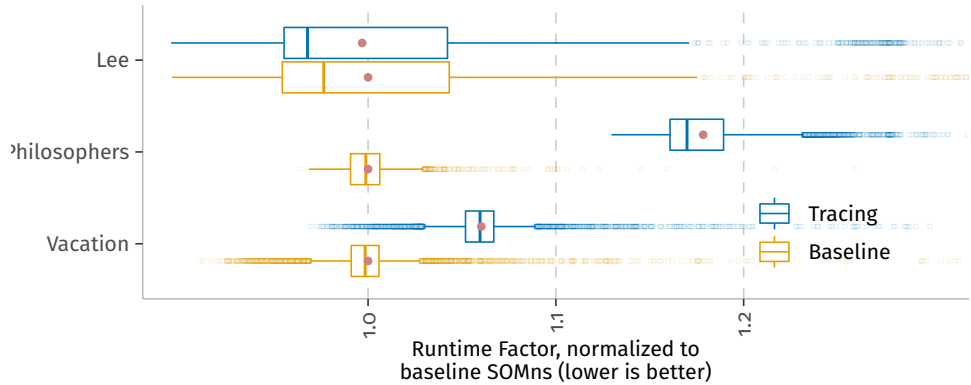

**Figure 10** Recording performance of our prototype for Threads & Locks compared to untraced execution showing results separately per benchmark. The average overhead is 7.85 % (min. −0.31 %, max. 17.84 %).

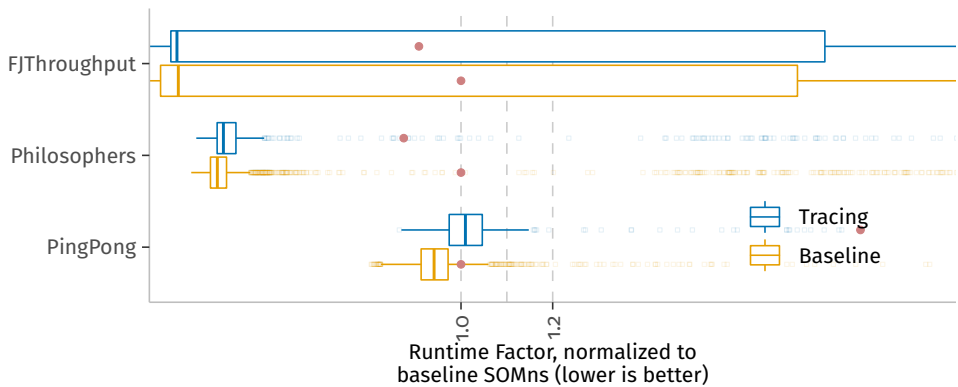

**Figure 11** Recording performance of our record & replay for a subset of the Savina benchmarks modified for CSP showing results per benchmark. The average overhead is 21.82 % (min. −12.5 %, max. 87.15 %).

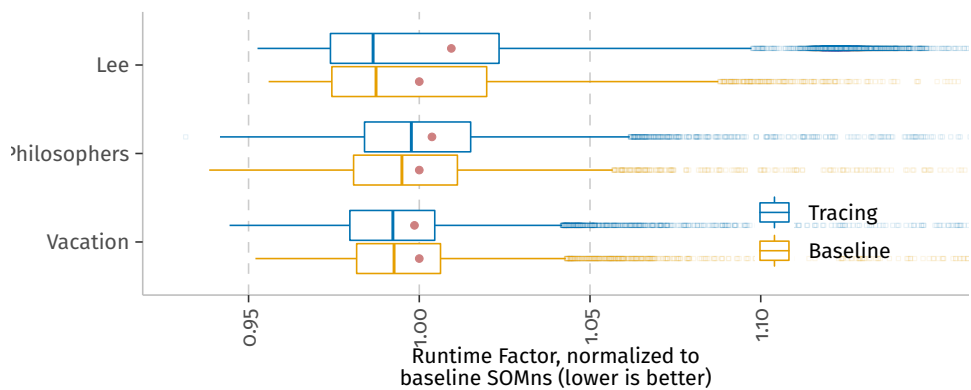

**Figure 12** Recording performance of our record & replay for the STM benchmarks, showing results per benchmark. Note that the STM implementation used in SOMns is not optimized and this, the tracing performance is not representative of optimized implementations.

### B.6 Trace Recording Performance based on Kind of Storage Device

The performance of trace recording is not only a function of instrumentation overhead, but we can determine and optimize this overhead without making assumptions for the storage device used for trace recoding. In figure 13, we show the difference in performance for recording a trace without writing it to storage, writing it to an solid state disk (SSD), and writing it to a hard disk drive (HDD) as measured on the Savina benchmarks.

We see that SSDs can be fast enough to not make a major difference. However, using a HDD instead of a SSD results on average in a 30.52 % overhead (min. −0.23 %, max. 107.67 %).

### B.7 Performance of Multithreaded Execution

As detailed in section 6.1, to be able to measure the overhead of tracing, we use generally single threaded execution. This means, effects like contention and context switching are avoided, and we can discern more clearly the impact of adding instructions into the program for trace recording.

However, to also give an impression of multithreaded execution, figure 14 shows the benchmark results on our octa-core AMD Ryzen 7 3700x without limiting SOMns to use a single thread, but allow it to use 8.

Compared to figure 4, has figure 14 a different scale, because the multithreaded execution experiences a much wider variation in measurements, since it is exposed to additional issues such has thread scheduling, context switching, cross-core caches, and contention, to name but a few.

The most important take away is that the impact of tracing is highly dependent on the specific nature of a program.

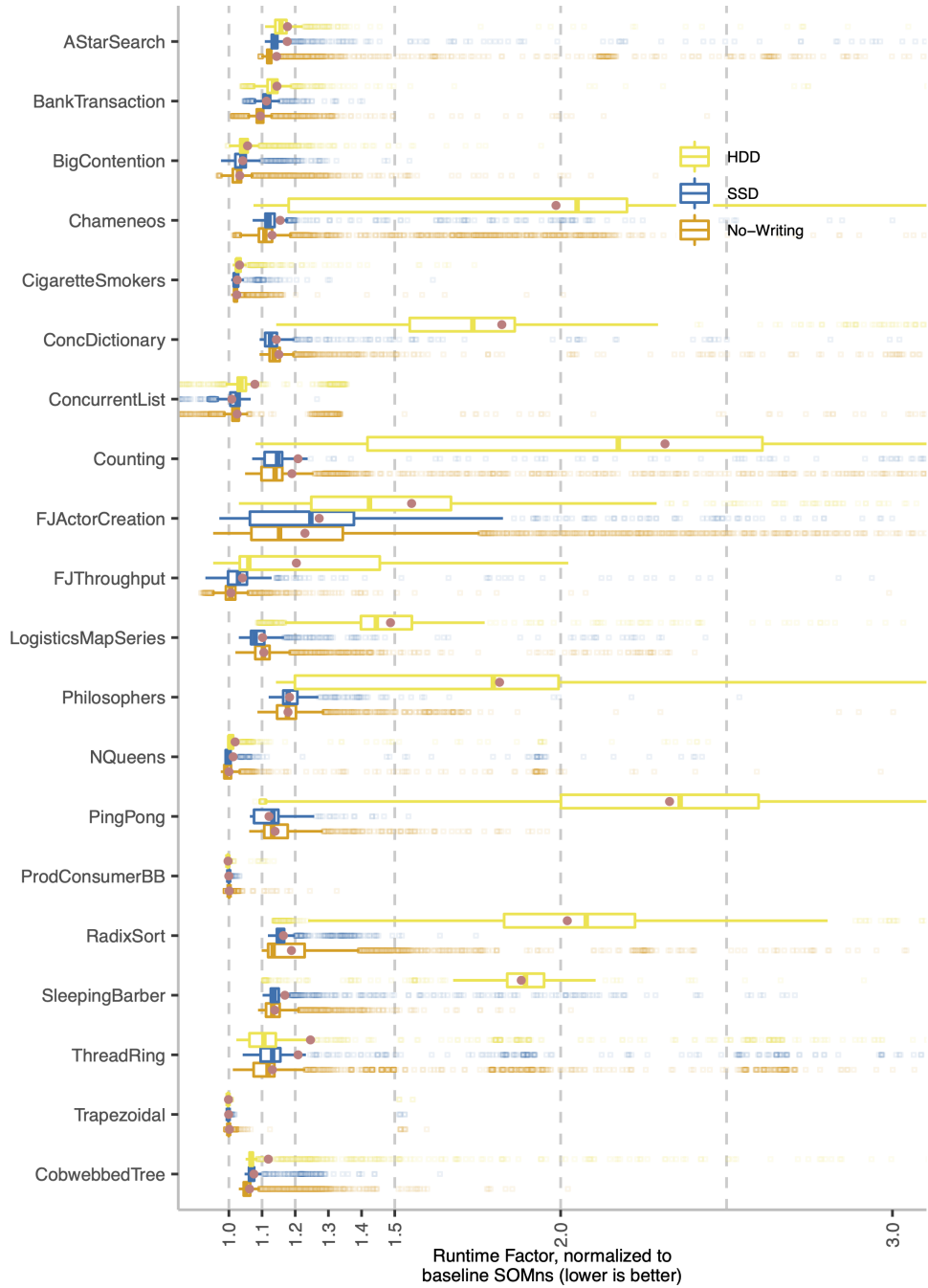

**Figure 13** Tracing performance of CEL actors for the Savina benchmark suite under different conditions. No-Writing represents the ideal case where writing to disk does not introduce overhead. The overhead introduced by writing traces to disk depends on both the characteristics of a benchmark and the available hardware.

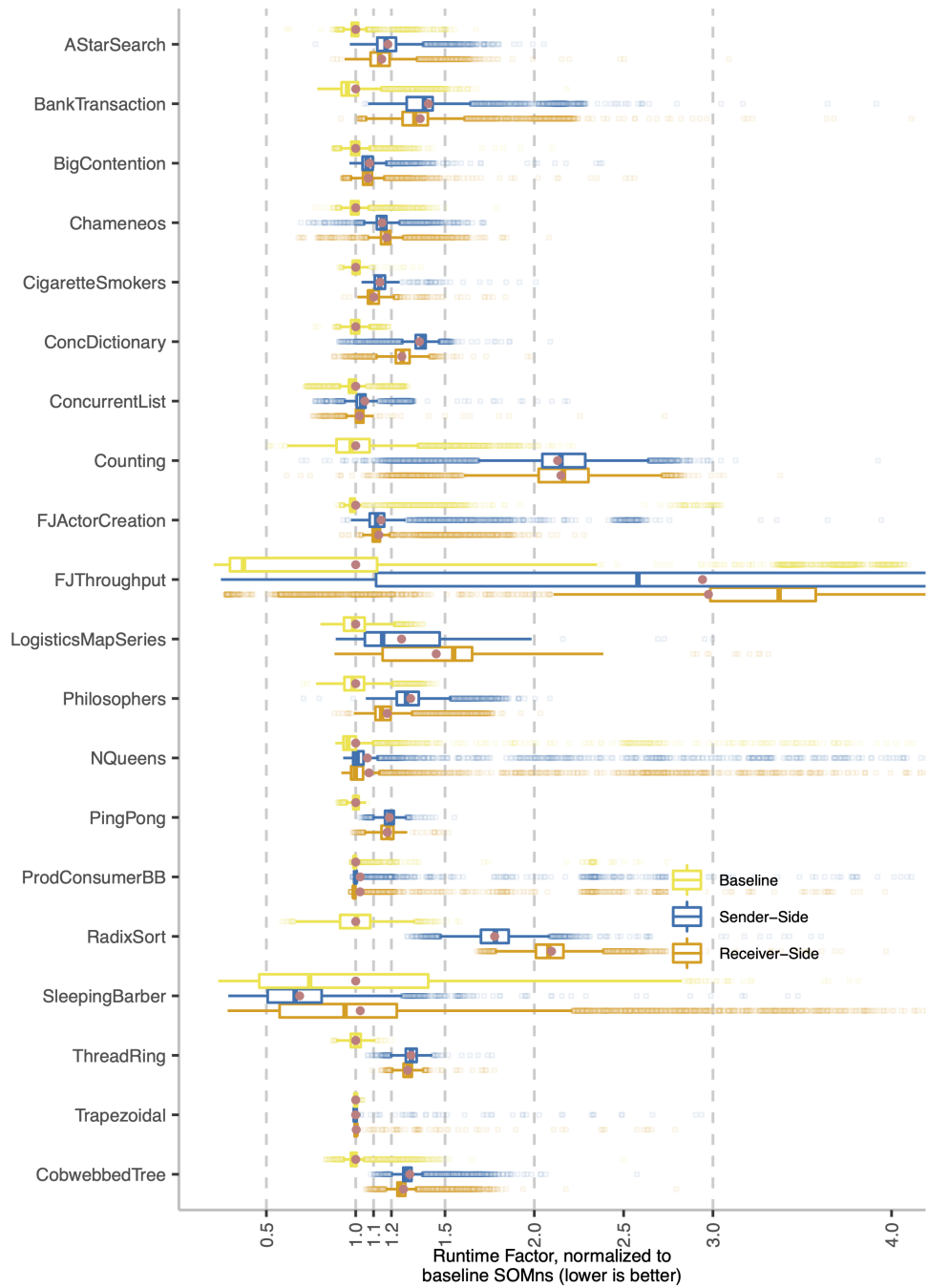

■ **Figure 14** Multithreaded performance of CEL actors for the Savina benchmark suite, the resulting traces are about the same size as those presented in table 2. The multi-threaded baseline run-time of some benchmarks is a multiple of singlethreaded execution.

■ **Table 2** Size of the traces recorded per benchmark iteration. The size factor is the trace size of our approach normalized to the Specialized approach. On average, our approach has a trace size overhead of 24 % (min. −34 %, max. 111 %), and is thus competitive with a special purpose approach.

|  | Mean MB/Iteration | | | Size Factor | |
| --- | --- | --- | --- | --- | --- |
|  | Sender Side | Receiver Side | Specialized | Sender Side | Receiver Side |
| AStarSearch | 5.47 | 2.67 | 2.59 | 2.11 | 1.03 |
| BankTransaction | 9.59 | 6.76 | 6.56 | 1.46 | 1.03 |
| BigContention | 9.12 | 7.20 | 7.20 | 1.27 | 1.00 |
| Chameneos | 8.38 | 6.71 | 6.71 | 1.25 | 1.00 |
| CigaretteSmokers | 1.52 | 1.49 | 1.46 | 1.04 | 1.02 |
| ConcDictionary | 7.39 | 7.71 | 7.51 | 0.98 | 1.03 |
| ConcurrentList | 0.57 | 0.59 | 0.57 | 0.99 | 1.04 |
| Counting | 6.75 | 10.80 | 10.20 | 0.66 | 1.06 |
| FJActorCreation | 7.40 | 3.76 | 3.68 | 2.01 | 1.02 |
| FJThroughput | 6.38 | 3.38 | 3.20 | 2.00 | 1.06 |
| LogisticsMapSeries | 20.94 | 16.75 | 16.25 | 1.29 | 1.03 |
| Philosophers | 11.74 | 12.77 | 12.39 | 0.95 | 1.03 |
| NQueens | 0.07 | 0.07 | 0.07 | 1.07 | 1.00 |
| PingPong | 7.35 | 6.96 | 6.84 | 1.07 | 1.02 |
| ProdConsumerBB | 0.05 | 0.04 | 0.04 | 1.25 | 1.00 |
| RadixSort | 10.12 | 15.30 | 14.45 | 0.70 | 1.06 |
| SleepingBarber | 36.19 | 35.70 | 34.94 | 1.04 | 1.02 |
| ThreadRing | 5.03 | 5.81 | 5.61 | 0.90 | 1.04 |
| Trapezoidal | 0.01 | 0.00 | 0.00 | NaN | NaN |
| CobwebbedTree | 14.55 | 9.84 | 9.64 | 1.51 | 1.02 |

**Capturing Nondeterminism for Concurrency Model Agnostic Record & Replay**

**About the authors**

**Dominik Aumayr** dominik.aumayr@jku.at

**Stefan Marr** s.marr@kent.ac.uk

**Sophie Kaleba** s.kaleba@kent.ac.uk

**Elisa Gonzalez Boix** egonzale@vub.be

**Hanspeter Mössenböck** hanspeter.moessenboeck@jku.at